\documentclass[twocolumn]{revtex4}
\usepackage{latexsym}
\usepackage{amsmath}
\usepackage{times}
\usepackage{amssymb}
\usepackage{fancyheadings}
\usepackage[T1]{fontenc}
\usepackage[utf8]{inputenc}
\usepackage{fancyhdr}
\usepackage{url}
\usepackage{hyperref}
\usepackage{color}

\usepackage{graphicx}

\usepackage{bm}

\begin{document}


\title{\bf Construction, sensitivity index, and synchronization speed of optimal networks}
\author{Jeremie Fish$^{1,4}$}
\author{Jie Sun$^{1,2,3,4}$}\thanks{Corresponding author. Email: sunj@clarkson.edu,  Tel.: +1 315 2682388,}

\affiliation{$^1$Department of Physics, Clarkson University, Potsdam, NY 13699, USA}
\affiliation{$^2$Department of Mathematics, Clarkson University, Potsdam, NY 13699, USA}
\affiliation{$^3$Department of Computer Science, Clarkson University, Potsdam, NY 13699, USA}
\affiliation{$^4$Clarkson Center for Complex Systems Science, Clarkson University, Potsdam, NY 13699, USA}
\date{\today}


\begin{abstract}
\noindent The stability (or instability) of synchronization is important in a number of real world systems, including the power grid, the human brain and biological cells. For identical synchronization, the synchronizability of a network, which can be measured by the range of coupling strength that admits stable synchronization, can be optimized for a given number of nodes and links. Depending on the geometric degeneracy of the Laplacian eigenvectors, optimal networks can be classified into different sensitivity levels, which we define as a network's sensitivity index. We introduce an efficient and explicit way to construct optimal networks of arbitrary size over a wide range of sensitivity and link densities. Using coupled chaotic oscillators, we study synchronization dynamics on optimal networks, showing that cospectral optimal networks can have drastically different speed of synchronization. Such difference in dynamical stability is found to be closely related to the different structural sensitivity of these networks: generally, networks with high sensitivity index are slower to synchronize, and, surprisingly, may not synchronize at all, despite being theoretically stable under linear stability analysis.

\vspace*{2ex}\noindent\textit{\bf Keywords}: Synchronization, Chaotic Oscillators, Optimal Networks, Sensitivity Index, Stability
\\[3pt]
\noindent\textit{\bf PACS}:  89.75.-k, 05.45.Xt
\\[3pt]
\noindent\textit{\bf MSC}: 34C15
\end{abstract}

\maketitle

\thispagestyle{fancy}


\section{Introduction}

Synchronization of network coupled systems has a wide range of real world applications \cite{strogatz2003,pikovsky2003,arenas2008}. This type of behavior has been observed in biological, chemical, mechanical and electrical systems, \cite{strogatz1993,camazine2003,tyson1973,field1972,nuno2011,blaabjerg2006,tadic2016} among others.
Stable synchronous states may be desirable, such as in the power grid where an asynchronous system may lead to cascading power failures \cite{meyers2013, nishikawa2015}. In other cases, synchronization can be undesirable, such as in the brains of Parkinson's patients, where excessive synchronization in certain parts of the brain  leads to reduced control over motor skills \cite{brown2003, hammond2007}. The study of synchronization and its stability has attracted extensive research efforts in the past two decades, including both theoretical and experimental investigations in many fields of science and engineering~\cite{pecora1998,terry1999,lu2002, wiley2006,sun2009a,sun2009b,pecora2014, menck2013,huddy2016,schroder2016,zhang2017,shirasaka2017}.

Focusing on network coupled dynamical systems, one may ask the question of how large is the range of coupling strength that guarantees stable synchronization, and further, what networks are optimal in the sense of maximizing such range. Recent research on addressing these important questions have yielded discovery of structural characteristics of optimal networks \cite{motter2005,chavez2005,donetti2005,nishikawa2006,nishikawa2006PRE, nishikawa2010,skardal2014}.
In particular, under the constraints of fixed number of nodes and edges, optimal networks are always {\it directed} networks except for the special case of a complete network where every pair of nodes is connected by a bidirectional edge. Interestingly, optimal networks that have identical number of nodes, edges, and range of stable coupling strengths can nevertheless yield significantly different synchronization profiles \cite{ravoori2011}.

In this paper, we introduce sensitivity index to quantify the structural sensitivity of optimal networks, enabling a finer distinction among optimal networks.
We adopt a generalized Krapivsky-Redner network growth model and show that by appropriate choice of the initial network, one obtains optimal networks. By controlling the redirection parameter, we found that the model produces optimal networks with a range of sensitivity indices, thus offering a first explicit and efficient construction of sensitive optimal networks.
We study synchronization speed and time using the dynamics of coupled chaotic oscillators on optimal networks. Our numerical simulations show that optimal networks that are more sensitive with respect to structural perturbations (thus having higher sensitivity index) tend to be slower in synchronization, and may sometimes not synchronize at all despite being deemed synchronizable under linear stability analysis.

\section{Stability of Network Synchronization}
We consider a widely used model for the synchronization of network coupled oscillators~\cite{pecora1998}. The dynamics of the individual oscillators are given by: 
\begin{equation}\label{eq:main}
\dot{x_i} = f_i(x_i) + \varepsilon\sum \limits_{j=1}^n A_{ij} h(x_j-x_i),~i=1,2,\dots,n.
\end{equation}
Here $f_i:\mathbb{R}^d\rightarrow\mathbb{R}^d$ represents the vector field of the isolated node dynamics, $h:\mathbb{R}^d\rightarrow\mathbb{R}^d$ is the coupling function, and $\varepsilon\in\mathbb{R}$ is the global coupling strength. Matrix $A=[A_{ij}]_{n\times n}$ is the {\it adjacency matrix} of the network, where $A_{ij}\neq0$ if and only if there is a directed link $j\rightarrow i$. The value of $A_{ij}$ can be interpreted as the relative weight of coupling from node $j$ to node $i$. 
The coupling received at each oscillator is scaled by a single parameter (global coupling strength). Such choice does not exclude the possibility of heterogeneous local couplings, the latter can be modeled by weighted edges in the network (non-binary entries in the adjacency matrix $A$).
Throughout the paper we focus on {\it simple} networks, where there is no self loop (thus $A_{ii}=0$ for every $i$) and the edges are unweighted ($A_{ij}\in\{0,1\}$). Furthermore, we assume that the network contains a directed spanning tree, that is, there exists a node $i_0$ such that for every other node $j\neq i_0$ there is a directed path of finite length $\ell$: $v_0\rightarrow v_1\dots\rightarrow v_\ell$, where $v_0=i_0$ and $v_\ell=j$.
Note that this ``spanning tree'' condition implies that the network is weakly connected without requiring the network to be strongly connected.

Assuming that the oscillators are identical ($f_i\equiv f,\forall i$) and coupling is diffusive ($h(0)= 0$), the network dynamics~\eqref{eq:main} admits synchronous trajectories that follow the isolated single-oscillator dynamics
\begin{equation}\label{eq:single}
	\dot{s} = f(s).
\end{equation}
The collection of synchronous trajectories are embedded in the synchronization manifold given by
\begin{equation}
\mathcal{M} = \{(x_1,\dots,x_n)|x_1=\dots=x_n\in\mathbf{R}^d\}.
\end{equation}

\subsection{Linear Stability of Identical Synchronization}
A practical question regarding synchronization is that of stability. That is, whether the system can maintain its synchronization against perturbations. If the initial perturbation to each node, denoted by $\delta_i\in\mathbf{R}^d$, is sufficiently small, one can linearize Eq.~\eqref{eq:main} around the synchronization trajectory, to obtain the so-called variational equations
\begin{equation}\label{eq:var}
	\dot{\delta_i} = Df(s) - \varepsilon Dh(0)\sum_{j=1}^{n}L_{ij}\delta_j,~i=1,\dots,n,
\end{equation}
where $D(\cdot)$ denotes taking Jacobian and the network {\it Laplacian} $L=[L_{ij}]_{n\times n}$ is defined as
\begin{equation}\label{eq:L}
L_{ij}=
\begin{cases}
-A_{ij}, & i\neq j\\
\left(\sum_{k\neq i}A_{ik}\right)-A_{ii}, &i=j.
\end{cases}
\end{equation}
Collecting the individual perturbation vectors into a single column vector $\bm{\delta}=[\delta_1^\top,\dots,\delta_n^\top]^\top$, the set of individual variational equations~\eqref{eq:var} can be reduced to a single high-dimensional variational equation
\begin{equation}\label{eq:var2}
	\bm{\dot{\delta}} = \left[I_n\otimes Df(s) - K L\otimes Dh(0)\right]\bm{\delta},
\end{equation}
where the symbol ``$\otimes$" denotes Kronecker product. Note that the variational equation~\eqref{eq:var2} is a linear time-dependent ODE (it becomes time-independent when the synchronous state $s$ is a fixed point), which describes the evolution of an initially ``small'' state perturbation over time.
If $\bm{\delta}\rightarrow0$ as $t\rightarrow\infty$, a small initial perturbation vanishes asymptotically and synchronization is linearly (or locally) stable; otherwise synchronization is not stable.

\subsection{Master Stability Analysis}
For a given system, one can in principle solve the full, high-dimensional variational equation~\eqref{eq:var2} to determine the stability of synchronization. However, to do so requires numerically integrating a system of potentially very high dimensions. Moreover, the variational equation itself does not directly reveal how the structure of a network might impact synchronization stability let alone designing networks that can optimize synchronization. To overcome these limitations, Pecora and Carroll proposed to decouple the high-dimensional variational equation~\eqref{eq:var2} into a set of low-dimensional, {\it master stability equations (MSEs)}~\cite{pecora1998}. Specifically, the starting point is to assume that the Laplacian matrix $L$ is diagonalizable, with
\begin{equation}\label{eq:Ldiag}
	L = V\Lambda V^{-1},
\end{equation}
where $\Lambda=\mbox{diag}{(\lambda_1,\lambda_2,\dots,\lambda_n)}$ is a diagonal matrix whose diagonal elements are the eigenvalues of $L$, ordered such that
\begin{equation}
	\operatorname{Re}{(\lambda_1)}\leq\operatorname{Re}{(\lambda_2)}\dots\leq\operatorname{Re}{(\lambda_n)},
\end{equation}
and the columns of $V$ are the corresponding eigenvectors. By definition, the row sums of $L$ are all zeros, implying that $\lambda_1=0$ (called the null eigenvalue) with associated (null) eigenvector 
$\bm{v}^{(1)}=[1,\dots,1]^\top$. Under the assumption of the network containing a directed spanning tree, all other eigenvalues of $L$ have positive real part, that is, $\operatorname{Re}(\lambda_i)>0$ for all $i>1$. Using the coordinate transformation
\begin{equation}
	\bm{\eta}=(I_d\otimes V^{-1})\bm{\delta},
\end{equation}
the high-dimensional variational equation~\eqref{eq:var2} decomposes into a set of $n$ low-dimensional variational equations,
\begin{equation}
	\dot{\eta}_i = [Df(s)-K\lambda_i Dh(0)]\eta_i,~i=1,2,\dots,n.
\end{equation}
The equation associated with the null eigenvalue $\lambda_1=0$ corresponds to perturbation mode parallel to the synchronization manifold whereas the others correspond to perturbations that are transverse to the synchronization manifold and therefore relevant to stability of synchronization. Given that these low-dimensional equations are identical except for the parameter $\lambda_i$, one can define a generic, one-parameter family of ODEs,
\begin{equation}\label{eq:mse}
	\dot{\eta} = [Df(s)-\alpha Dh(0)]\eta,
\end{equation}
referred to as master stability equations~\cite{pecora1998}. For given $f$ and $h$, the master stability function (MSF)
\begin{equation}
	\Omega:\mathbb{C}\rightarrow\mathbb{R}
\end{equation}
is defined as follows. For $\alpha\in\mathbb{C}$, the value of $\Omega(\alpha)$ is the maximal Lyapunov exponent of~\eqref{eq:mse} associated with the dynamics~\eqref{eq:single}, which gives the asymptotic rate of convergence (or divergence) with respect to the state $\eta=0$.

\begin{figure}[htbp]
\flushleft
\includegraphics[width = 0.5\textwidth]{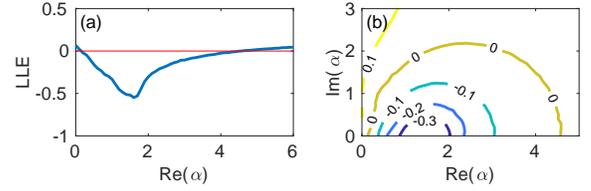}
\caption{Master stability function (MSF) $\Omega(\alpha)$ of coupled R\"{o}ssler oscillators. The oscillator model $f$ and coupling function $h$ are given in Eq.~\eqref{eq:rossler} and Eq.~\eqref{eq:rosslercoupling}, respectively. 
(a) Values of $\Omega(\alpha)$ for real-valued $\alpha\in[0,6]$. The horizontal line marks the stability threshold $\Omega=0$.   
(b) Values of $\Omega(\alpha)$ for complex-valued $\alpha$ where $\operatorname{Re}(\alpha)\in[0,5]$ and $\operatorname{Im}(\alpha)\in[0,3]$, shown as a contour plot. \label{fig1:rosslermsf}}
\end{figure}

As an example, in Fig.~\ref{fig1:rosslermsf} we show the MSF computed for coupled R\"{o}ssler oscillators. The individual (isolated) dynamics of the R\"{o}ssler oscillators are specified by
\begin{equation}\label{eq:rossler}
	f(x)=f([x^{(1)},x^{(2)},x^{(3)}]^\top)=
	\begin{cases}
	-x^{(2)}-x^{(3)},\\
	x^{(1)}+ax^{(2)},\\
	b+(x^{(1)}-c) x^{(3)},
	\end{cases}
\end{equation}
with parameters $a=b=0.2$, and $c=7$. The coupling function $h:\mathbb{R}^3\rightarrow\mathbb{R}^3$ is given by
\begin{equation}\label{eq:rosslercoupling}
	h(\Delta{x}) = H\Delta{x},
\end{equation}
where $H=\mbox{diag}([1,0,0])$. Thus, coupling is through the $x^{(1)}$ component of each oscillator.
Computation of the Lyapunov exponents are done using the MATLAB implementation of Wolf's algorithm as developed in~\cite{wolf1985}.

\subsection{Region of Coupling for Stable Synchronization}
For a given oscillator model and coupling function that specifies $f$ and $h$ in Eq.~\eqref{eq:main}, the master stability function defines a master stability region in the complex plane:
\begin{equation}
	S(f,h) = \{\alpha:\Omega(\alpha)<0\}\subset\mathbb{C}
\end{equation}
for which perturbations that are transversal to the synchronization manifold asymptotically decay to zero. 
Depending on the type of dynamics and coupling, the master stability region can be either infinite or finite, either convex or nonconvex, and either connected or disconnected \cite{huang2009}.
Synchronization of a particular network is stable if the scaled eigenvalues $\varepsilon\lambda_i$ are all within the master stability region for all nontrivial eigenvalues $\{\lambda_i\}_{i=2}^{n}$ of the network Laplacian matrix, that is, if the following condition holds
\begin{equation}
	\{\varepsilon\lambda_i:i=2,\dots,n\}\subset S(f,h).
\end{equation}
Furthermore, for a given network, the condition implies that the set of all possible coupling strengths for stable synchronization is given by
\begin{equation}
	E(L)=\{\varepsilon:\varepsilon\lambda_i(L)\in S(f,h),~\forall i=2,\dots,n\},
\end{equation}
which we refer to as the region of coupling for stable synchronization, or simply stability region of coupling.

\section{Optimal Network Structure for Identical Synchronization}
From the master stability analysis, it becomes clear that the Laplacian eigenvalues play a central role in determining the synchronizability of a network. Generally, the more tightly clustered the nontrivial eigenvalues are, the more flexible the network is to admit stable synchronization because it is easier to select coupling strength for the scaled eigenvalues to be contained inside the master stability region(s). To quantify synchronizability independent of a particular oscillator model, one can consider the normalized spread of nontrivial Laplacian eigenvalues, defined as~\cite{nishikawa2010}
\begin{equation}
	\sigma^2(L) = \frac{1}{d^2(n-1)}\sum_{i=2}^{n}|\lambda_i-\bar\lambda|^2,
\end{equation}
where $d=\frac{1}{n}\sum_{i}L_{ii}$ is the average degree of the network, and $\bar\lambda=\frac{1}{n-1}\sum_{i=2}^{n}\lambda_i$ is the average of the nontrivial Laplacian eigenvalues. Generally, the smaller the spread $\sigma^2$ is, the more synchronizable the network is.

\subsection{Structural Characterization of Optimal Networks}
For fixed number of nodes $n$ a natural question is, which network(s) yield the minimal spread $\sigma^2$ and therefore optimize synchronizability? This question was addressed in depth in Ref.~\cite{nishikawa2010}, giving rise to the following results depending on the number of directed edges $m$:
\begin{itemize}
\item If $m=k(n-1)$ for some $k\in\mathbb{N}$, there exists networks with $\sigma=0$, that is,
\begin{equation}
	\lambda_2(L)=\dots=\lambda_n(L).
\end{equation}
Since $\sigma\geq0$ by definition, networks that achieve $\sigma=0$ are termed {\it optimal} and minimize the spread of nontrivial Laplacian eigenvalues.
\item If $m=k(n-1)+r$ for some $k,r\in\mathbb{N}$ and $1\leq r<n-1$, all networks with $n$ nodes and $m$ edges yield $\sigma>0$ and therefore cannot be (strictly) optimal. 
\end{itemize}

What are the structural properties of networks that are optimal (i.e., with $\sigma=0$)? Perhaps the most basic property is that, 
unless the network is a {\it complete graph} (whose adjacency matrix is given by $A_{ij}=1$ for all $i\neq j$ and $A_{ii}=0$ for every $i$),
the network can be optimal only if the Laplacian matrix is not diagonalizable~\cite{nishikawa2006}.
Therefore, except for the trivial scenario of a complete network (only possible when $m=n(n-1)/2$), optimal networks are always {\it directed} because the Laplacian matrix of an undirected network is symmetric and thus diagonalizable. Fundamentally, a network being optimal requires its nontrivial Laplacian eigenvalues to be completely degenerate ($\lambda_2=\dots=\lambda_n$).

\begin{figure*}[htbp]
\includegraphics[width=0.95\textwidth]{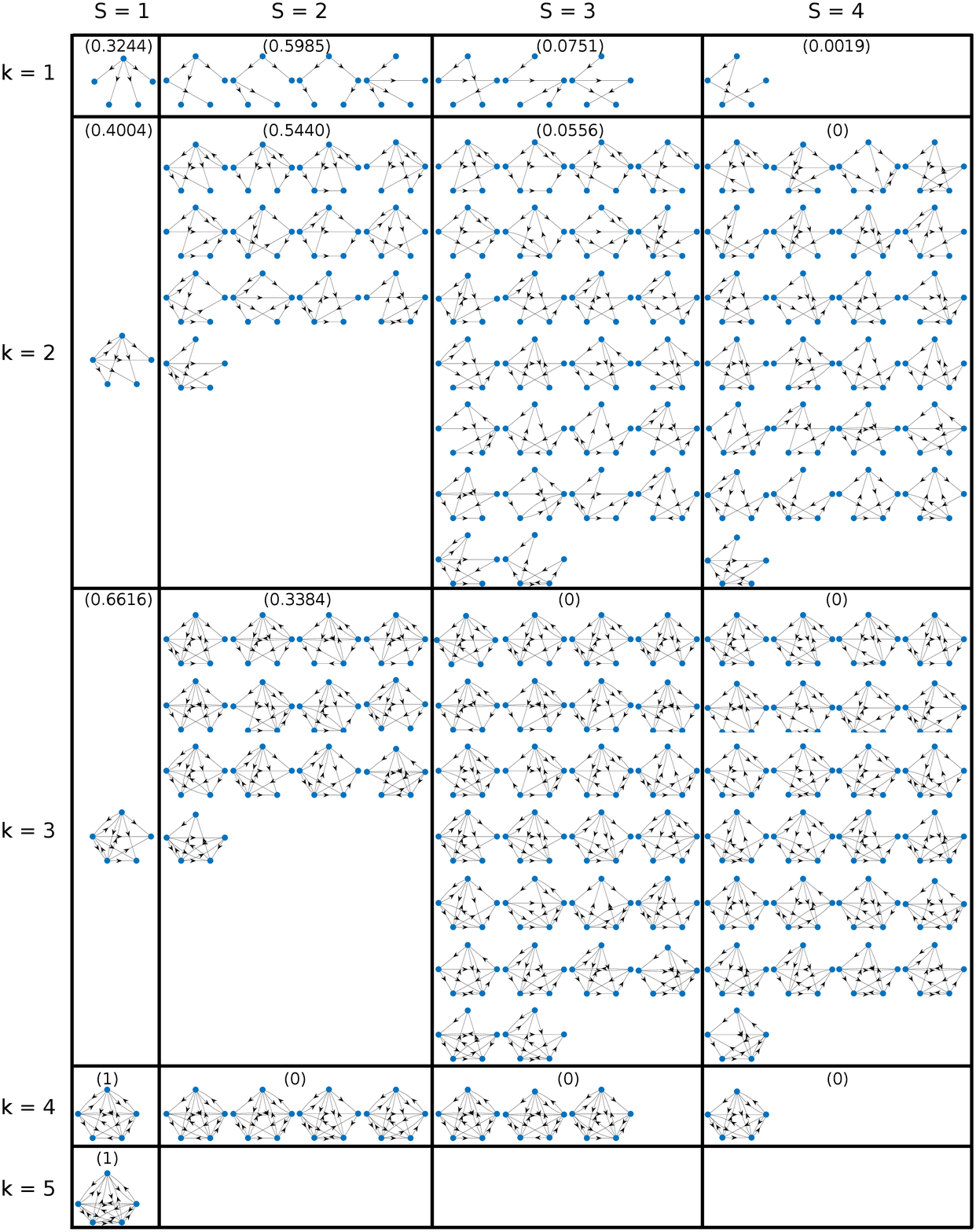}
\caption{Non-sensitive and sensitive optimal networks with $n=5$. At the top of each cell the probability listed corresponds to the probability of drawing networks of the given sensitivity index $S$ at the given $k$ value using the generalized Krapivsky-Redner model.}
\label{fig:optimaln5}
\end{figure*}

Below we give a complete characterization of optimal networks up to $n=5$:
\begin{itemize}
\item $n=2$: the possible number of directed edges that yield optimal networks are $m=1$ and $m=2$. In both cases there is only 1 network (up to isomorphism) and the network is optimal.
\item $n=3$: the possible number of directed edges that yield optimal networks are $m=2$, $m=4$, and $m=6$. The $m=6$ case corresponds to the complete network, which is optimal. When $m=2$, there are two non-isomorphic optimal networks: one network is the directed linear chain $1\rightarrow2\rightarrow3$, corresponding to $A_{21}=A_{32}=1$ and $A_{ij}=0$ otherwise;  and the other network is the directed star with two links $1\rightarrow2$ and $1\rightarrow3$, thus $A_{21}=A_{31}=1$ and $A_{ij}=0$ otherwise. For $m=4$, there are also only two non-isomorphic optimal networks, which can be obtained by taking the {\it complement} of the optimal networks of $m=2$. Note that there is a simple rule relating the adjacency matrix $A$ of a graph and that of its complement, denoted by $A^{c}$: $A^{c}_{ij}=(1-\delta_{ij})(1-A_{ij})$.
\item $n=4$: there are $18$ non-isomorphic optimal networks (4 with $m=3$, 9 with $m=6$, 4 with $m=9$, and 1 with $m=12$), which were reported in the Supplementary Material of Ref.~\cite{ravoori2011}.
\item $n=5$: through exhaustive search~\footnote{
We first obtain the set of all non-isomorphic unweighted directed graphs of size $n = 5$ using the ``digraph" function in Sage (also commonly referred to as SageMath). Then, optimal networks are selected as the graphs whose nontrivial Laplacian eigenvalues are completely degenerate (that is, $\lambda_2=\dots=\lambda_n>0$), where the eigenvalues are computed symbolically in Sage.}, we found 149 non-isomorphic optimal networks. These networks, together with the corresponding value of $k$ (which determines $m$ as $m=k(n-1)$), are shown in Fig.~\ref{fig:optimaln5}. 
\end{itemize}

\section{Sensitivity Index of Optimal Networks}
Given the abundance of optimal networks, and their apparent structural differences (see for example Fig.~\ref{fig:optimaln5}), a natural question is: do these structural differences bear anything relevant to the dynamics on the networks, in particular synchronization properties? This important question was addressed in Ref.~\cite{ravoori2011}, where lab experiments using optoelectronic oscillators show that optimal networks can exhibit qualitatively and drastically different synchronization properties. For example, the time it takes for a particular network to reach synchronization can vary greatly among different optimal networks, even though these networks have the same number of nodes, same number of edges, and identical Laplacian eigenvalues.

Optimal networks can be classified into {\it sensitive} and nonsensitive, according to the geometric degeneracy of the nontrivial Laplacian eigenvalues~\cite{ravoori2011}. In fact, the {\it algebraic multiplicity} (or degeneracy) of the nontrivial Laplacian eigenvalues of any optimal network is $n-1$, since optimal implies that $\lambda_2=\dots=\lambda_n=\lambda_*(>0)$. However, the {\it geometric degeneracy} of these eigenvalues can vary greatly from one optimal network to another. In particular, when the geometric multiplicity of $\lambda_*$ is $n-1$, there are $n-1$ linearly independent eigenvectors associated with $\lambda_*$, and the optimal network is termed nonsensitive; otherwise the optimal network is called sensitive. One reason for these terms is that, when considering small, random structural perturbations to the network, so that $L\rightarrow L+\varepsilon\Delta{L}$, the change of the Laplacian eigenvalues $\lambda\rightarrow\lambda+g(\varepsilon)$ following an approximate scaling formula~\cite{nishikawa2016},
\begin{equation}\label{eq:eigscaling}
	g(\varepsilon)\sim\varepsilon^{1/\beta}.
\end{equation}
When $\beta>1$, the derivative of $g$ with respect to $\varepsilon$ diverges and the function $g$ becomes non-differentiable at $\varepsilon=0$, and thus the Laplacian eigenvalues exhibit sensitive dependence on the network structure and edge weights; on the other hand, when $\beta=1$, it follows that $g'(\varepsilon)$ is a constant and therefore the Laplacian eigenvalues show no sensitive dependence on the network.
The quantitative nature of the parameter $\beta$ provides basis for a finer classification of the sensitive networks. 
In Ref.~\cite{nishikawa2016}, the scaling parameter $\beta$ is shown to relate to the size of the Jordan blocks associated with the network Laplacian. 
In particular, consider the Jordan normal form of the Laplacian matrix $L$ of an optimal network,
\begin{equation}
J(L) = 
\begin{pmatrix}
  0 & 	 &  &  \\
     & J_1 &  &  \\
    &   & \ddots &   \\
   &  &  & J_q
 \end{pmatrix}_{n\times n},
\end{equation}
where each $J_i$ ($i=1,\dots,q$) is a Jordan block
\begin{equation}
J_i=
\begin{pmatrix}
  \lambda &1 	 &  &  \\
     & \lambda &  \ddots &  \\
    &   & \ddots &  1 \\
   &  &  & \lambda
 \end{pmatrix}_{n_i\times n_i},
\end{equation}
arranged in a way such that
\begin{equation}
	n_1\geq\dots\geq n_q\geq1.
\end{equation}
Note that $\sum_{i=1}^{q}n_i=n-1$, and $1\leq q\leq n-1$.
It follows that the scaling parameter $\beta\approx n_1$ except for some special types of perturbations and network structures~\cite{nishikawa2016}.
Consequently, we define the {\it sensitivity index} of an optimal network to be the value of $n_1$. For the special case of $n_1=1$ (thus $q=n-1$), the network has the smallest sensitivity index $1$ and is called nonsensitive; otherwise, whenever $n_1>1$, the network is sensitive.

\section{Generation of optimal networks with controlled sensitivity}
Having developed the sensitivity index as a measure for the structural sensitivity of optimal networks, we now wish to address the problem of generating optimal networks with prescribed sensitivity index.

We first review the generation of nonsensitive optimal networks (sensitivity index equals 1), as previously reported~\cite{nishikawa2006}. Give $n$ and $m=k(n-1)$, the idea is to form the network as a union of $k$ directed stars. In particular, define entries of the adjacency matrix as
\begin{equation}
A_{ij}=
\begin{cases}
1-\delta_{ij},~j\leq k,\\
0,~j>k.
\end{cases}
\end{equation}
It follows that the eigenvalues of the Laplacian matrix are $\{0,k,\dots,k\}$, and there are $n-1$ linearly independent eigenvectors associated with the (repeated) eigenvalue $\lambda=k$, thus $n_1=1$ and the network is nonsensitive. This construction, however, cannot be readily extended to produce optimal networks with sensitivity index greater than one.

Next, we note that any directed tree network is always optimal, with the sensitivity index equalling the depth of the tree. Thus, if $m=n-1$, one can construct optimal networks with arbitrary sensitivity index by controlling the depth of the corresponding directed tree.

\begin{figure*}[htbp]
\includegraphics[width=0.95\textwidth]{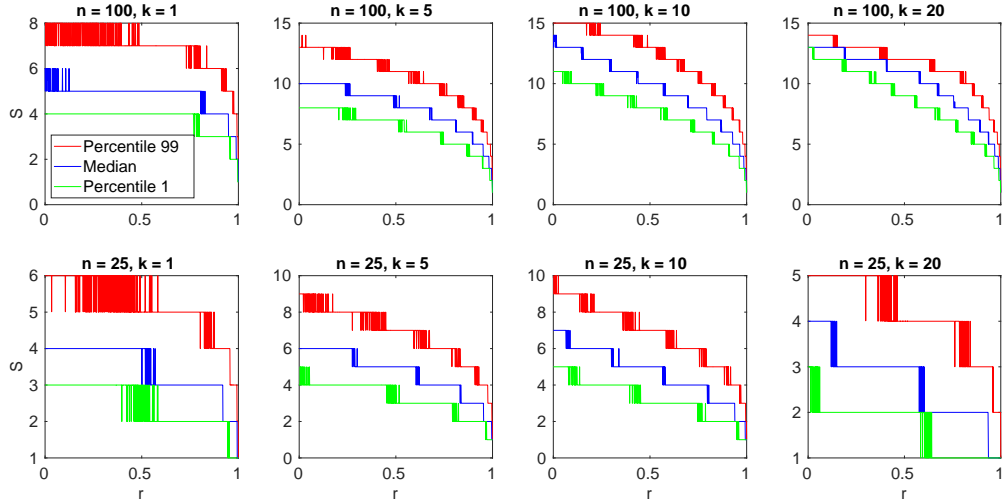}
\caption{Sensitivity index of optimal networks constructed by a generalized KR model (see text for details). Here for each parameter combination $(n,k,r)$, we computed and show the median value, $1$st percentile, as well as the $99$th percentile of the sensitivity index, all estimated from $1000$ independent realizations of the model.}
\label{fig:sind}
\end{figure*}

Finally, for the general case of $m=k(n-1)$ ($k\in\mathbb{N}$), there has been no previous studies on how to construct optimal networks with controlled sensitivity. We here offer a method to accomplish this task, by considering a modified version of the Krapivsky-Redner (KR) growth model~\cite{rozenfeld2004}, with a single {\it redirection} parameter $r\in[0,1]$. The starting point is a complete network of $k$ nodes (also called a $k$-clique). Then, the additional $n-k$ nodes are added to the existing network one at a time. In each time step $i$ ($i=1,2,\dots,n-k$), a node (indexed as $i$) joins the existing network and makes $k$ directed links. Each one of these $k$ links are formed as follows. First, a node $j$ is chosen at random among the existing nodes in the network ($j\in\{1,2,\dots,i-1\}$). Then, with probability $1-r$, the link $j\rightarrow i$ is formed; otherwise, with probability $r$, another node $\ell$ is chosen randomly among the directed neighbors of node $j$, that is, from the set $\mathcal{N}_j=\{\ell:A_{j\ell}=1\}$, and the link $\ell\rightarrow i$ is formed instead. At the end of the process, we obtain an optimal network of $n$ nodes and $m=k(n-1)$ directed links (the network Laplacian has an eigenvalue $\lambda=k$ with algebraic multiplicity $n-1$).
With a given set of parameters ($n$ and $m=k(n-1)$)  this generalized KR model enables us to produce optimal networks, whose sensitivity can be varied by controlling the redirection parameter $r$. In particular, the choice of $r=1$ always produces an optimal nonsensitive network. Generally, the smaller the value of $r$ is, the more sensitive the network tends to be.
In Fig.~\ref{fig:sind} we show the sensitivity index of optimal networks generated using the proposed generalized KR model, for several choices of parameter combinations. For fixed number of nodes $n$, we found that the method produces larger range of sensitivity index under larger value of $m$ when the network is sparse. Keep increasing the value of $m$ can eventually decrease the range of sensitive index (for the extreme case of $m=n(n-1)/2$, the network is always nonsensitive, being a complete graph).

\section{Synchronization time and speed of optimal networks}
Our final set of numerical experiments aims at exploring the impact of sensitivity index on the synchronization speed and time of optimal networks. To ensure a fair comparison, we construct optimal networks of the same number of nodes $n$ and same number of directed links $m$. Here $n=100$ and $m=5(n-1)$. By using different values of $r\in\{0.01,0.25,0.75,0.95,0.99,1\}$, we obtain six networks with distinct sensitivity index, $S=1,3,5,7,9,11$. We consider coupled R\"{o}ssler oscillators as described by Eqs.~\eqref{eq:rossler} and~\eqref{eq:rosslercoupling} on these networks. Since the Laplacian eigenvalues of these six networks are the same, with $\lambda_1=0$ and $\lambda_2=\dots=\lambda_n=5$ for every network, an extended version of the master stability analysis (see Ref.~\cite{nishikawa2006}) implies that synchronization is stable for all these networks so long as the coupling strength $\varepsilon\in(0.02464, 0.9326)$. We selected four coupling strengths in this interval: $\varepsilon=0.82,0.85,0.87,0.89$, for our numerical experiments. To quantify synchronization, we define the synchronization error for the network state $(x_1(t),x_2(t),\dots,x_n(t))$ as
\begin{equation}
	e(t) = \max_{i,j}\|x_j(t)-x_i(t)\|.
\end{equation}

\begin{figure*}
\includegraphics[width=0.95\textwidth]{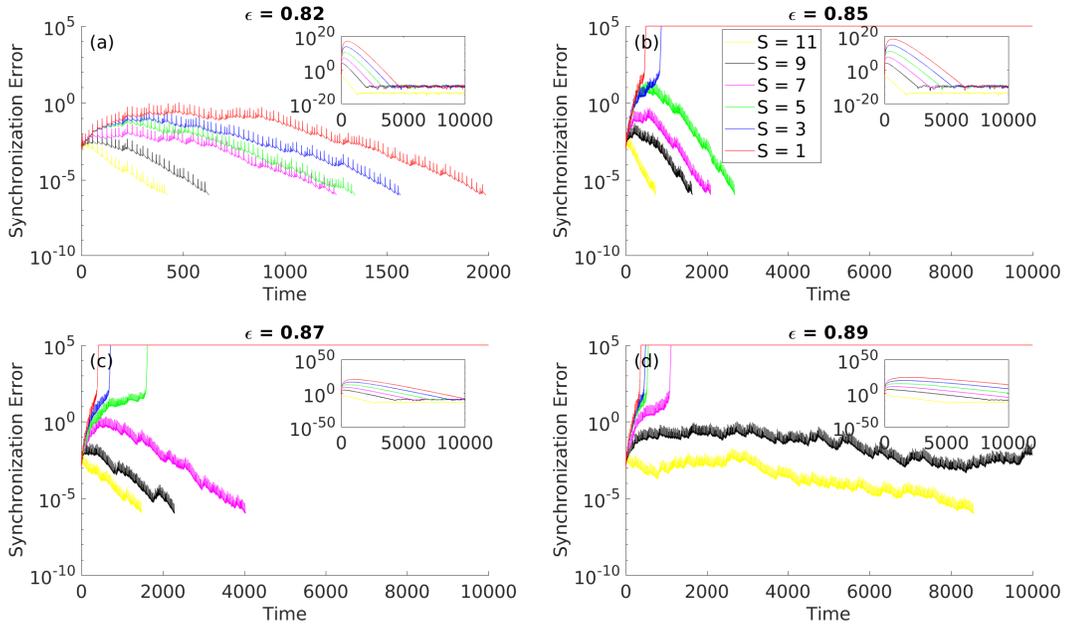}
\caption{Synchronization error as functions of time for optimal networks of different sensitivity index.
In (a) the coupling strength $\varepsilon = 0.82$ and all of the networks synchronize, in (b) $\varepsilon = 0.85$ and despite the fact that this is in a predicted linear stability regime the two most sensitive network desynchronize while the other networks synchronize, in (c) $\varepsilon = 0.87$: some of the networks synchronize, others do not and  (d) $\varepsilon = 0.89$ all of the networks except for the 2 least sensitive networks synchronize, despite being in the linearly stable region.}
\label{fig:syncerror}
\end{figure*}

In Fig. ~\ref{fig:syncerror}, we show that for a random initial condition where the initial state of each oscillator is independently drawn from the uniform distribution in $[10-0.005,10+0.005]^3$, while keeping the same initial condition for all six networks, the dynamics of these networks exhibit quite different behavior. For the coupling strength $\varepsilon=0.82$, all networks synchronize, with the less sensitive networks (those with smaller sensitivity index) synchronize faster. Interestingly, and perhaps surprisingly, for the other coupling strenghts, some of the more sensitive networks in fact lose synchronization, despite the fact that synchronization is linearly stable for all networks. This loss of synchronization cannot be interpreted by a linearization-based (local stability) theory, since the equivalent linear system would always converge to the (global) stable state despite initial transient growth. The actual nonlinear system, however, can have a local basin of synchronization. If the initial transient growth is too fast, as seen in some of the very sensitive networks, it can ``kicks'' the state of the system outside of the basin, and thus prevent the system from returning to eventual synchronization.

\begin{figure*}[htbp]
\includegraphics[width=0.9\textwidth]{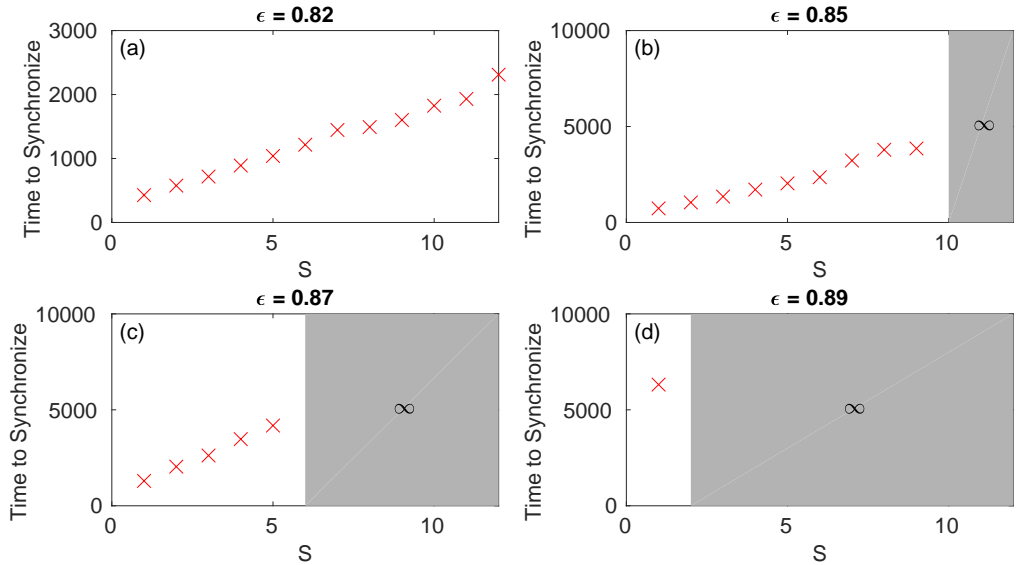}
\caption{The time to synchronization in sensitive and nonsensitive optimal networks. 
In (a) the coupling strength $\varepsilon= 0.82$  all of the networks synchronize for all initial conditions, in (b) $\varepsilon = 0.85$ the three most sensitive networks have at least one initial condition for which they do not synchronize. (c) $\varepsilon = 0.87$ the 5 least sensitive networks synchronize for all initial conditions  (d) $\varepsilon = 0.89$ only the least sensitive network synchronizes.}
\label{fig:synctime}
\end{figure*}

To investigate the dependence of synchronization error and speed on network sensitivity in a more systematically manner, we generated, in addition to the six networks considered in Fig.~\ref{fig:syncerror}, six additional networks with sensitivity index $S=2,4,\dots,12$, using values of $r\in\{0.01,0.2,0.65,0.85,0.98,0.999\}$. We numerically simulate the coupled R\"{o}ssler oscillators on each of the twelve networks, for $250$ random initial conditions, where the initial state of each oscillator is independently chosen uniformly at random from the set $[10-0.005,10+0.005]^3$. Note that the same initial conditions are used for all the twelve networks. We then record the time it takes for each network under each initial condition to synchronize, defined as the first time at which the synchronization error $e(t)$ is less than $10^{-6}$. If such state is never reached, the time to synchronization is considered $\infty$. In Fig.~\ref{fig:synctime} we plot the average time to synchronize for the networks as functions of the networks' sensitivity index, for several coupling strengths. We found that it takes longer time to synchronize for optimal networks that are more sensitive (higher sensitivity index). Under certain coupling strength, the most sensitive networks can in fact lose synchronize, despite predicted otherwise by a linearization-based local stability theory such as the master stability analysis.

\section{Conclusion}
In this work we quantify sensitivity of optimal networks using the size of the largest Jordan block of a network's Laplacian matrix, called the network's sensitivity index. 
We developed a computationally efficient approach to construct optimal networks of arbitrary size using a generalization of the Krapivsky-Redner network growth model, and show that the approach can explicitly produce optimal networks of a range of sensitivity indices by controlling a single redirection parameter. Using coupled R\"{o}ssler oscillators, we study the speed and time of synchronization on optimal networks that are identical in all means except for their sensitivity index. We found that networks with higher sensitivity index generally take a longer amount of time to reach synchronization. Interestingly, a sensitive optimal network sometimes does not synchronize at all, which we attribute to the more profound transient growth of perturbations in these networks (comparing to less sensitive ones) as well as the non-global nature of their basins of synchronization.

\end{document}